\documentclass[reprint,aps,pra]{revtex4-1}
\usepackage{amsmath}
\usepackage{bm}
\usepackage{graphicx}
\usepackage{epstopdf}
\usepackage{subfigure}

\newcommand{\bra}[1]{\langle #1 | \,}
\newcommand{\ket}[1]{\, | #1 \rangle}

\begin{document}

\title{Non-destructive photon detection using a single rare earth ion coupled to a photonic cavity}

\author{Chris O'Brien}
\affiliation{Institute for Quantum Science and Technology and Department of Physics and Astronomy, University of Calgary, Calgary AB T2N 1N4, Canada}
\email{cobrien.physics@gmail.com}
\author{Tian Zhong}
\affiliation{T.J. Watson Laboratory of Applied Physics, California Insitute of Technology, 1200 E California Blvd, Pasadena, CA, 91125, USA}
\author{Andrei Faraon}
\affiliation{T.J. Watson Laboratory of Applied Physics, California Insitute of Technology, 1200 E California Blvd, Pasadena, CA, 91125, USA}
\author{Christoph Simon}
\affiliation{Institute for Quantum Science and Technology and Department of Physics and Astronomy, University of Calgary, Calgary AB T2N 1N4, Canada
}

\begin{abstract}
We study the possibility of using single rare-earth ions coupled to a photonic cavity with high cooperativity for performing non-destructive measurements
of photons, which would be useful for global quantum networks and photonic quantum computing. We calculate the achievable fidelity as a function of the parameters of the rare-earth ion and photonic cavity, which include
the ion's optical and spin dephasing rates, the cavity linewidth, the single photon coupling to the cavity, and the detection efficiency.
We suggest a promising experimental realization using current state of the art technology in Nd:YVO$_4$.
\end{abstract}

\maketitle

\section{Introduction}
The ability to detect photonic qubits non-destructively would be very useful for many quantum information applications, including long-distance quantum communication \cite{Repreview,satellite} and photonic quantum computing \cite{PQCtheory, PQCexp,OBrien_2003}. One approach for non-destructive measurement is to use a single atom or ion that is coupled with high cooperativity to an optical cavity \cite{CPHASEexp}. In particular, Ref. \cite{Duan_2004} suggested realizing a quantum controlled phase-flip
(CPHASE) gate between the photon and the ion based on the fact that, depending on the state of the ion, the photon would either be reflected unchanged or with a $\pi$-phase shift.
The resulting entanglement between the photon and the ion can be used to detect the photon through readout of the ion's state. These ideas have recently been realized in a series of impressive experiments with single trapped atoms inside free-space high-finesse cavities \cite{Rempe_2013,Rempe_2014,Rempe_2015, Rempe_2016}.

For more robust and scalable technologies, it would be useful to be able to implement similar protocols in the solid state. A single rare-earth ion (REI) doped into a crystal is very similar to an optically trapped single atom, 
when the crystal is cooled to cryogenic temperatures in order to avoid dephasing via coupling to phonons.  Rare-earth doped crystals have been successfully used for optical quantum memories \cite{GisinQM, KrollQM, LongdellQM},
and have been suggested for scalable quantum computing \cite{Kroll_2007}.
A scheme for performing non-destructive measurements utilizing an ensemble of rare-earth ions coupled to a bulk crystalline waveguide has recently been suggested \cite{Sinclair_2015}.
It is also now possible to observe single rare-earth ions in bulk crystal \cite{Kolesov_2012, Rogge_2013, Sellars_2013, Sandoghdar_2014, Sandoghdar_2015}, and to map between ion-spins and a photon's polarization \cite{Kolesov_2013}.
There has recently been success in coupling Nd ions doped into yttrium orthosilicate (YSO) \cite{Faraon_2015} and yttrium orthovanadate (YVO) \cite{Faraon_2016} crystals with photonic crystal cavities that were fabricated out of bulk crystal.
The advantage of using rare-earth ions compared to other solid state emitters like nitrogen-vacancy centers in diamond or semiconductor quantum dots is that they combine narrow inhomogeneous broadening, low spectral diffusion, close to transform-limited optical line-widths and spin states with a long coherence time.

Using rare-earth ions doped into photonic cavities to realize CPHASE gates was first suggested by \cite{Longdell_2009}. Here we perform an in-depth analysis of this idea with a focus on the implementation of non-destructive photon detection, including a detailed scheme and an accounting for the likely fidelity. 
Rare-earth ions coupled to nano-photonic resonators will enable an on-chip platform where single ions act like optically addressable single quantum bits that can be interfaced via photons, with the possibility for on-chip photon storage into optical quantum memories made from the same atomic species. 

The paper is organized as follows. In section II we introduce the protocol for creating the conditional phase shift. In section III we explain how this setup can be used for quantum non-destructive measurements. In section IV we discuss how the state of the ion can be read out. 
In section V we calculate the fidelity of the non-destructive measurement as a function of the rare-earth ion and photonic cavity parameters,
including the dephasing of the ion's optical and spin transitions, the linewidth of the cavity, the single photon coupling between the ion and the cavity, as well as the probability of successful read out. In section VI we discuss a specific implementation of
the protocol in Nd:YVO$_4$ crystals. In section VII we give some concluding remarks.
\section{Conditional Phase Shift}
\label{sect:cphase}
Consider a single rare-earth ion doped directly into a photonic crystal cavity, where one side is partially transparent and the other end is perfectly reflecting.
The incoming photon will interact with the
ion-cavity system. If the ion is strongly coupled to the cavity, the photon will reflect off the cavity with a phase that depends
on the state of the ion. If the ion is not coupled to the cavity, the photon will enter the cavity and receive a $\pi$-phase shift.
If the ion is coupled to the cavity, the cavity will not be impedance matched with the photon, so the photon will
reflect off the cavity without entering and will not have a phase shift. Thus the reflected photon gains a phase that is dependent
on whether the ion is in a state that interacts with the cavity, which creates a CPHASE gate.

If we assume the input field is weak enough to have a low probability to excite the ion,
we can write down a set of quantum Langevin equations \cite{Longdell_2009}:
\begin{align}
& \dot a(t) = (-\kappa -i\delta) a(t) +gs(t) - \sqrt{2\kappa} a_{in} (t), \label{eq:a} \\
& \dot s(t) = -g a(t) + (-\gamma/2 -i\delta -i\Delta) s(t). \label{eq:s}
\end{align}
Here $\kappa$ is the decay rate of the cavity, $\delta$ is the detuning of the incoming photon from the cavity, $\gamma$ is the decoherence rate of the ion, g is the single photon coupling between the REI and the cavity, $a(t)$ is the photon excitation amplitude, $s(t)$ is the atomic excitation amplitude, and $a_{in}$ is the amplitude of the photon incident on the cavity.
We will see that the probability of a single photon entering the cavity and exciting the atom is inversely proportional to the single ion cooperativity $C = g^2/(\kappa \gamma)$ which we take as large and thus justify
our assumption that the atom remains in its ground state.

These equations can be solved under the assumption that the input field has a narrow frequency range
with respect to the dynamics of the atom-cavity system such that we can perform adiabatic elimination, i.e. $\dot a(t) = \dot s (t) = 0$.
Assuming there is no initial excitation of the atom then the output photon can be expressed as  a function of the input photon.
\begin{align}
a_{out} (\delta) = \frac{g^2 +(i\delta + i\Delta + \gamma/2)(i\delta - \kappa)}{g^2 + (i\delta +i\Delta +\gamma/2)(i\delta +\kappa)} a_{in} \label{eq:out}
\end{align}
This expression covers two cases, when the ion is in resonance with the cavity, we can take $\Delta = 0$, which will give us the case where the photon
does not enter the cavity. Then when we do not want the ion interacting with the cavity, we put the ion into a metastable state that is far detuned from the cavity with $\Delta \gg 0$, which will allow the photon to enter the cavity and receive a $\pi$-phase shift.
In general, for a REI, both of the ground states will interact with an upper transition,
for the same polarization of applied light, which is not necessarily the case for a trapped ion, which is why we need to consider the case
of the non-resonant transition rather than just assuming that one state of the ion does not interact with the cavity at all as \cite{Duan_2004, Longdell_2009} assumes.

Both transitions being allowed is one difference between the protocol in trapped atoms and REI.
Another difference is that REI tend to have weaker dipole moments than those of trapped atoms.
For trapped atoms that are strongly
coupled to cavities it is typical to be in the ``good cavity'' regime where  $g \gg \kappa \gg  \gamma$, but for the REI-cavity system this is unlikely. Since the ion-cavity coupling
is usually weaker due to smaller dipole moments.
The REI-cavity system
is instead in what is called the ``bad cavity'' regime \cite{Longdell_2009} where $\kappa \gg g \gg  \gamma$ and yet the single photon cooperativity is still high with $C\gg1$.

\begin{figure}
\centering
\includegraphics[width=8cm]{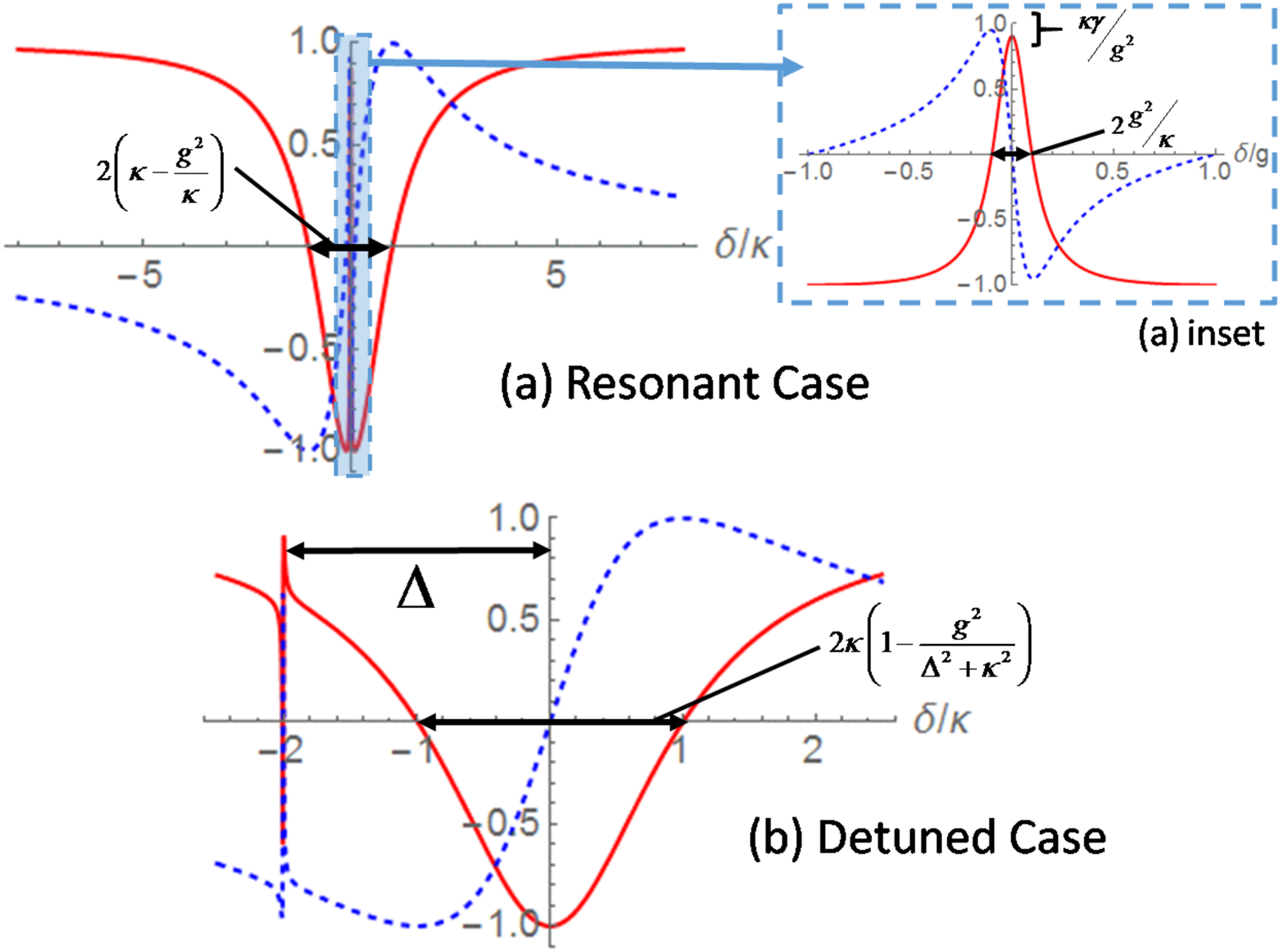}
\caption{(color online) Plot of the real (solid, red) and imaginary (dashed, blue) parts of $a_{out}/ a_{in}$ given by Eq.(\ref{eq:out}), in the bad cavity regime. The plot lists the frequency widths and amplitudes that are analytically derived in Sect.\ref{sect:cphase}. The parameters are normalized to $g = 1$, with $\kappa = 10g$, and $\gamma = 0.01g$. (a) When the atom is in the state that is resonant with the cavity,
plotted in units of $\delta/\kappa$. Inset: zooming in on near resonance plotting in units of $\delta/g$. This is the case that limits the bandwidth of an input photon. (b) We plot the ratio $a_{out}/a_{in}$ given by Eq.(\ref{eq:out}) in units of $\delta/\kappa$, for when the atom is in the state that is far-detuned from the cavity, with the previous values and a detuning of $\Delta = 20g$,
we also assume that $\tilde{g} = g$.
}	\label{fig:cphase}
\end{figure}

In order to understand the details of the CPHASE protocol, we develop a clear analytic picture, as shown in Fig.\ref{fig:cphase}.
Using the decaying-dressed state analysis from \cite{Petr,OBrienRb}, we can analyze Eq.(\ref{eq:out}) to get simple analytic expression for the amount of the phase shifts and the bandwidth over which they occur.
In the bad cavity limit, we can expand Eq.(\ref{eq:out}) into partial fractions. If the ion
is in resonance such that $\Delta=0$, then
\begin{align}
\frac{a_{out}}{ a_{in}} (\delta) = 1 +& \frac{2i\kappa\left(1-g^2/\kappa^2\right)}{\delta -i\kappa +ig^2/\kappa}- \frac{2ig^2/\kappa}{\delta - ig^2/\kappa-i\gamma/2}. \label{eq:aratio}
\end{align}
The coupling between the atom and cavity creates a broad region with a HWHM of $\kappa-g^2/\kappa$ where the photon enters the cavity and gets a $\pi$-phase shift with a narrow central feature with HWHM of $g^2/\kappa+\gamma/2$ where the
interaction with the atom stops the photon from entering the cavity as shown in Fig. \ref{fig:cphase}(a).
For near resonance $\delta \approx 0$, the ratio of output and input is close to unity when the single photon cooperativity is high
\begin{align}
\frac{a _{out}}{a_{in}} (0, \Delta=0) = 1  - \frac{\kappa \gamma}{g^2} \label{eq:dzbc}
\end{align}
such that an incoming photon is reflected with no phase change. 
The reflectivity is not exactly unity because a small amount of the photon enters the cavity and is scattered by the atom.

Now we consider when the ion is in the far-detuned state such that $\kappa,\Delta\gg g\gg \gamma$
\begin{align}
&\frac{a_{out}}{a_{in}} (\delta) = 1 + \frac{2i\kappa \left( 1- \frac{\tilde{g}^2}{(\Delta+i\kappa)^2}\right) }{\delta -\frac{\Delta \tilde{g}^2}{\Delta^2+\kappa^2} -i\kappa \left( 1 - \frac{\tilde{g}^2}{\Delta^2+\kappa^2}\right) } \nonumber \\
&+\frac{2i\kappa \frac{\tilde{g}^2}{(\Delta+i\kappa)^2}}{\delta+\Delta \left( 1+\frac{\tilde{g}^2}{(\Delta^2+\kappa^2)} \right) -i\gamma/2-i\kappa\frac{\tilde{g}^2}{(\Delta^2+\kappa^2)}}.
\end{align}
Since the detuned transition may be weaker due to the partial selection rules of the REI we label the cavity-ion coupling for this transition as $\tilde{g}$
to distinguish it.
If the atom is in its far-detuned state, then the atomic resonance is far-detuned from the photon frequency with $\Delta \gg \gamma$, then there is the cavity interaction centered at $\delta = \Delta\tilde{g}^2/(\Delta^2+\kappa^2)$ and a Fano resonance $\delta = -\Delta(1+\tilde{g}^2/(\Delta^2+\kappa^2))$ as shown in Fig. \ref{fig:cphase}(b). The first term has a HWHM of $\kappa(1 -\tilde{g}^2/(\Delta^2+\kappa^2))$ and is due to interaction with the cavity, where the photon enters the bad cavity and then leaves with a $\pi$-phase shift. The second feature is too far detuned to interact directly with the photon.
The ratio for a photon with frequency near the cavity resonance $\delta \approx 0$ is then
\begin{align}
\frac{a_{out}}{a_{in}} (0) = -1 -2i \frac{\tilde{g}^2}{\kappa \Delta}. \label{eq:dpbc}
\end{align}
The imaginary term is due to residual far-detuned interaction with the ion, which causes a small phase shift of the photon.

Eq.(\ref{eq:out}) was derived in the adiabatic limit, dropping the time derivatives of the field and atom.
Now consider the case where the photon that reflects off the cavity has a finite bandwidth. Taking a Gaussian pulse
with pulse duration HWHM is $T_p$ centered at $\delta = 0$, Eq.(\ref{eq:out}) can be averaged over the bandwidth of the pulse under the assumption that $1/T_p > g^2/\kappa$.
This updates Eq.(\ref{eq:dzbc}) to
\begin{align}
\frac{a _{out}}{a_{in}} (\Delta=0) = \left(1  - \frac{\kappa \gamma}{g^2} \right) e^{-\frac{\kappa \sqrt{\log 2}}{\pi T_p g^2}}, \label{eq:dzbc2}
\end{align}
such that it now applied to a finite pulse.

\section{Non-destructive photon measurement}
\label{sect:qnd}
This setup can be used as a method of entangling photons and single rare-earth ions, for use in quantum computing or for non-destructive photon detection.
The basic idea is similar to \cite{Rempe_2014}.
To generate entanglement we first initialize the ion in a superposition of the ground states, then an incoming photon will be put in a superposition state with a $\pi$-phase shift entangled with the ion.
Then by performing a rotation and measurement on the ion state, we can read out whether there was an incoming photon.

First, prepare the ion in a superposition of the two ground states
\begin{align}
\ket{\phi _a} = \frac{1}{\sqrt{2}} \left( \ket{0}_a+ \ket{1}_a \right)
\end{align}
where $\ket{1}_a$ is the ground state of the far-detuned transition between $\ket{1}_a$ and $\ket{e}_a$ cavity, and $\ket{0}_a$ is the ground state of the resonant transition between $\ket{0}_a$ and $\ket{e}_a$.
with the cavity. For a REI, this superposition can be created by using a pair of externally applied far-detuned Raman pulses to drive the system
into this state. For a single REI this is more straightforward then for a crystal with a high density of ions, because with a high density of ions,
it is required to use extensive hole-burning \cite{transfer}, to isolate those spins with a particular frequency.
The prepared superposition state can live only as long as the coherence remains, so once we turn off the external pulses we
will only have a limited time to perform the non-destructive photon measurement depending on the decoherence rate of the spin transition.

The photon that reflects off of the cavity can be in a superposition state of a single photon$\ket{1}_p$ and the state with no
photon is $\ket{0}_p$ with an arbitrary phase such that our photon state is
\begin{align}
\ket{\phi _p } = c_0 \ket{0}_p + c_1 \ket{1}_p
\end{align}
which as we showed in Sect.\ref{sect:cphase}, will give a $\pi$-phase
shift to the state where there is both a photon present and the atom is in $\ket{0}_a$.
This leads to a combined entangled state:
\begin{align}
\ket{\phi} = &\frac{1}{\sqrt{2}} c_0 \ket{0}_p (\ket{0}_a +  \ket{1}_a)   \nonumber \\
+&\frac{1}{\sqrt{2}} c_1 \ket{1}_p (\ket{1}_a -  \ket{0}_a)
\end{align}
Then performing a $\pi$/2 rotation of the ion state which once again can be performed with external pulses
such that
\begin{align}
R_a (\pi/2) \ket{0}_a &= \frac{1}{\sqrt{2}} \ket{1}_a - \frac{1}{\sqrt{2}} \ket{0}_a , \\
R_a (\pi/2) \ket{1}_a &= \frac{1}{\sqrt{2}} \ket{1}_a + \frac{1}{\sqrt{2}} \ket{0}_a.
\end{align}
Then the ideal entangled state is
\begin{align}
\ket{\phi_{ideal}} = & c_0 \ket{0}_p \ket{1}_a - c_1 \ket{1}_p \ket{0}_a \label{eq:ideal}
\end{align}
Finally a measurement is made on the atom in the population basis.

This completes our non-destructive measurement, since now by detecting the photons emitted into the cavity, we know if a photon
reflected off of the cavity. The photon reflected off the cavity may have a phase shift, but is otherwise unchanged by the process.
Normally, detecting a photon necessitates its destruction.
With a unheralded time bin qubit, a non-destructive photon measurement can be made on both bins in order to know there is a photon in one without
destroying the qubit \cite{Sinclair_2015}. With a heralded time bin photon, we could entangle the REI with the time bin qubit by limiting our QND measurement
to one of the time bins.

\section{Read out}
\label{sect:readout}

In order to identify that a photon has reflected off of the cavity it is necessary to detect that the ion was in the state $\ket{0}_a$.
This can be accomplished by optically pumping this level to the excited state with a narrow band laser and then detecting the
fluorescence.
For a REI the probability of detection is  limited by the branching ratios for fluorescence from the excited state to lower energy levels. This is greatly improved by interaction with the cavity, due to the Purcell effect.
The Purcell effect is due to the density of states for the cavity being much larger than the density of states for free space.
The rate of emission into the cavity is enhanced by the Purcell factor defined as:
\begin{align}
F_P = \frac{3}{4\pi^2}\left(  \frac{\lambda}{n} \right)^3 \left( \frac{Q}{V} \right), \label{eq:Purcell}
\end{align}
where $\lambda$ is the wavelength of the cavity, $n$ is the refractive index of the crystal, $Q$ is the quality factor of the cavity, and $V$ is the mode volume of the cavity.
For a two level atom, the Purcell factor is related to the single ion cooperativity through the ratio of the radiative line width $\gamma_r$ to the total linewidth $\gamma$ through $F_P = (\gamma/\gamma_r) C$, and thus $F_P$ is always larger than $C$.
Thus, for $C\gg1$ there is a much
higher chance that fluorescence will be into the cavity mode, rather than free space. If a single rare earth ion is strongly coupled to a high quality photonic crystal cavity, it is possible to reach Purcell factors greater than $F_P = 1,000$.
Now in a multi-level atom, there will be multiple channels for fluorescence, the probability in the bare ion to fluoresce in the desired channel is given by
the branching ratio $\beta$ for that transition. This probability is enhanced by interaction with the cavity such that
\begin{align}
p_{cav} = \frac{F_P \beta}{1-\beta+F_P \beta }
\end{align}
Then for example if we had a branching ratio of $\beta = 15\%$ and $F_P = 1000$, this would give the probability of fluorescence into the cavity as $p_{cav} = 0.994$.

Utilizing preferential emission into the cavity, means the photons emitted into the cavity must be detected. So sometime after the time-bin photon reflects
off the cavity, the optical path should be switched such that any future photons emitted from the cavity can be detected by a single photon detector.
Then the detection is limited by the efficiency of the single photon detector, which we will assume is $p_{det} = 0.9$.

When $p_{cav}>p_{det}$, the best way to read out the atomic state is to
drive the cavity transition itself ($\sigma$-polarized) and rely on the Purcell effect to preferentially fluoresce into the cavity mode
in order to have a cycling transition such that if the ion is in the proper state, many photons are emitted into the cavity.
If the ion does not fluoresce into the cavity, the photon is lost and the population cycling ends, which means this method has a maximum
efficiency of $p_{cav}$.

One issue with this method is there is a possibility that photons from the pump will be scattered into the cavity mode. 
This means that, besides
detector dark-counts, scattering will also lead to false-positives.
In this case, it is necessary to detect some minimum number of photons $n_M$, in order to discriminate against false positives.
Then the detection efficiency can be written as a sum over the number of photons created in the cavity,
with a factor giving the probability of $n$ photons being emitted and a factor determining the probability of detecting at least $n_M$
photons when $n$ photons are in the cavity
\begin{align}
\eta _{det} = \sum _{n=1} ^{\infty} p_{cav}^n(1 -p_{cav}) \sum _{k=n_M} ^n \binom{n}{k} p_{det}^k (1-p_{det})^{n-k}. \label{eq:detect}
\end{align}
For example, if we choose to detect $n_M = 2$ photons, to try to reduce dark counts, with $p_{cav} = 0.994$ and $p_{det} = 0.9$, then
the detection efficiency will be $\eta_{det} = 0.987$.
If it is determined that more photons are needed to get a signal above the background
of detector dark-counts and scattered light, then this efficiency does not decrease much, for $n_M = 4$, the efficiency is only a little lower $\eta_{det} = 0.974$.
Thus for the rest of the paper, we will assume this read-out method.
\begin{figure}
\includegraphics[width=6cm]{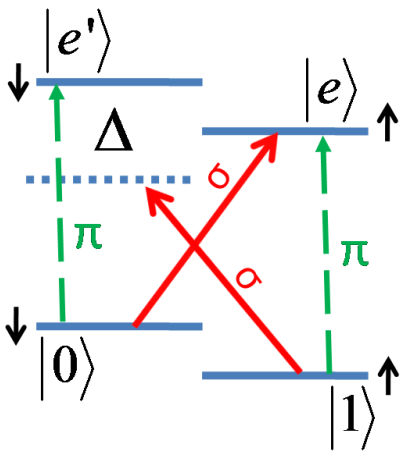}
\caption{(color online) Energy level diagram of 879.7nm transition in Nd:YVO$_4$. Showing the qubit states, and how the ion interacts with
$\sigma$-polarized light, i.e. electric field polarized parallel to the crystal axis.
Here the $\ket{0} \rightarrow \ket{e}$ transition is resonant with the cavity.
Both optical
transitions have the similar $g$ \cite{NDYVO}. In the presence of a 300mT magnetic field, applied at a 45\% degree angle with respect to the
crystal axis, the detuning for the non-resonant transition is $\Delta = 2\pi \cdot 9$GHz. } \label{fig:Ndlevel}
\end{figure}
%

The entire detection process must be completed before the state $\ket{0}$, relaxes to the ground state $\ket{1}$, but
the spin relaxation time $T_1$, is quite long for rare-earth ions kept below 7K, on the order of 100ms. This gives plenty of time to complete the read-out process.
Another concern is the possibility of false positives due to accidentally driving state $\ket{1}$ to emit a photon into the cavity.
Since the Purcell effect guarantees a high probability that a photon will be emitted into the cavity if the off-transition
is driven to the alternate excited state labeled $\ket{e'}$ in Fig.\ref{fig:Ndlevel}, we just need to calculate the probability to excite the
far-detuned transition.
The pump laser needs to have a Rabi frequency $\Omega = \tilde{\mu} E_p/\hbar$ that is large enough to achieve Rabi flopping with a significant
amount of the population reaching the excited state in order to have fast read-out. Here $\tilde{\mu}$ is the dipole moment of the driven transition, $E_p$ is the pump electric field, and $\hbar$ is Planck's constant divided by $2\pi$.  At the same time a larger $\Omega$ leads to quicker read-out but
also leads to a higher chance of driving the off-transition which may lead to a false positive.
A good compromise is to take $\Omega \geq \gamma$, but of similar magnitude.
Any lower $\Omega$ will lead to lower excited state population, which slows down the emission while the probability of exciting the off transition is equal to:
\begin{align}
p_{off} = \frac{|\Omega|^2}{\Delta^2} n_{cyc} \frac{\tilde{g^2}}{g^2} > \frac{\gamma^2}{\Delta^2} n_{cyc} \frac{\tilde{g^2}}{g^2}  \label{eq:off}
\end{align}
where $n_{cyc}$ is the average number of cycles that the driven transition goes through which for $p_{det} = 0.994$ is $n_{cyc} = 116$, and there is a factor $\tilde{g^2}/g^2$
to account for the possibility that the off-resonant transition is weaker than the resonant transition. We will estimate $p_{off}$
for our different schemes in Sect.\ref{sect:implement}, but in general it can be kept small enough to not hamper the fidelity.
At the same time, the selection rules for driving transitions in the REI are not perfect, such that sigma polarized light may still drive a predominately $\pi$-polarized transition, therefore read out may also lead to a false positive
due to population in $\ket{1}$ being driven to $\ket{e}$, then emitting into the cavity. This probability is similar to that given by Eq.(\ref{eq:off}), now with $\tilde{g}$ being the reduced interaction due to the polarization mismatch and $\Delta$
is just the energy difference between $\ket{0}$ and $\ket{1}$. Therefore, this probability is also low and can be safely neglected.

There are a few other ways to detect the atomic state.
One approach to spin selective detection is to pump the ground state into a higher level that has a fast non-radiative decay to our excited state, which then will preferentially fluoresce back to the ground state, as  demonstrated in Pr:YSO \cite{Sandoghdar_2014}.
This method is too slow and does not strongly  discriminate between the spin states.
Another approach, which is ideal when $p_{det} > p_{cav}$, is to drive a $\pi$-polarization transition of the REI to cycle population back and forth between state $\ket{1}$ and the excited state $\ket{e}$ as shown in Fig.\ref{fig:Ndlevel}.
Since an ion in $\ket{1}$ will cycle until it emits a single photon into the
cavity, the detection efficiency is only limited by the detector efficiency $\eta_{det} = p_{det}$.
Another approach is to use the phase shift present in the CPHASE gate, to detect if the ion is excited by reflecting a weak coherent beam off of the cavity and then measuring the phase shift.
The last approach is to utilize the change in reflectivity of the cavity when an ion is coupled to it, by reflecting a weak coherent pulse off the cavity and measuring the transmission as analyzed in \cite{Waks_2016}.
These last two techniques can have close to perfect read-out in a single-pass with the use of many photons. But for the current scheme, the number of photons must be limited to prevent exciting the ion.

\section{Fidelity}
\label{sect:fidelity}
In order to calculate the fidelity of the non-destructive photon measurement, we will work through the entire process.
In order to consider decoherence, this analysis is performed on a mixed state, using the density matrix formalism.
To simplify this analysis we will consider only the case when a single photon is present, which leads to a $2\times2$ atomic density matrix.
The presence of  a photon is the worse case for the fidelity as in the vacuum case the photon does not interact directly with the cavity,
so this assumption is justified.
The process starts with the ion in the ground state $\ket{1}_a$.
The next step is to rotate the ion into a superposition state $(\ket{0}_a + \ket{1}_a)/\sqrt{2}$ by applying a $\pi/2$ rotation through
the application of external fields. If this rotation is not perfect, then the rotation would be
at an angle $\pi/2 + \phi_P$ where $\phi_P$ is some small angle deviation.
Then the $\rho_a$ becomes $R _{\pi/2+\phi_P} \rho_a R _{\pi/2+\phi_P}^\dag$.
Now the ion will undergo dephasing, if we assume this is pure dephasing and not spin flipping, then this is handled by introducing
a dephasing rate $\gamma_{gs}$.
This dephasing continues for the entire time that the atom remains in the superposition state
which we will assume is a time period $T_{sp}$.
This period is at least as long as the time-bin photon, but in practice may need to be longer.
Then the density matrix is
\begin{align}
\rho_a &=  \left( \begin{array}{cc}
\frac{1}{2}-\frac{\phi_P}{2} & (\frac{1}{2}-\frac{\phi_P^2}{4})e^{-\gamma_{gs} T_{sp}} \\
(\frac{1}{2}-\frac{\phi_P^2}{4})e^{-\gamma_{gs} T_{sp}} & \frac{1}{2}+\frac{\phi_P}{2} \end{array} \right). \label{eq:rhoa}
\end{align}
Now consider the case that a photon reflects off the cavity then from Eq.(\ref{eq:dzbc2}) and Eq.(\ref{eq:dpbc}),
\begin{widetext}
the new density matrix is
\begin{align}
\rho_a &=  \left( \begin{array}{cc}
\frac{1}{2}(1-\phi_P)\left( 1 - \frac{\kappa  \gamma}{g^2} \right)^2 e^{-2\frac{\kappa \sqrt{\log 2}}{\pi T_p g^2}} & -\frac{1}{2}(1-\frac{\phi_P^2}{2})(1-\frac{\kappa \gamma}{g^2})(1-2i\frac{\tilde{g}^2}{\kappa \Delta})e^{-\gamma_{gs} T_{sp}}e^{-\frac{\kappa \sqrt{\log 2}}{\pi T_p g^2}} \\
 -\frac{1}{2}(1-\frac{\phi_P^2}{2})(1-\frac{\kappa \gamma}{g^2})(1+2i\frac{\tilde{g}^2}{\kappa \Delta})e^{-\gamma_{gs} T_{sp}}e^{-\frac{\kappa \sqrt{\log 2}}{\pi T_p g^2}} & \frac{1}{2}(1+\phi_P) | 1 + 2i \frac{\tilde{g}^2}{\kappa \Delta} |^2 \end{array} \right).
\end{align}
The  state must be rotated again by $\pi/2$ for read out. Assuming a small error
in creating the phase shift $\phi_R$ such that we rotate through $\pi/2 + \phi_R$. If we assume each correction is small and keep only the first order
terms then the density matrix becomes
\begin{align}
\rho_a &=  \left( \begin{array}{cc}
1-\frac{\kappa \gamma}{g^2}-\frac{\kappa \sqrt{\log 2}}{\pi T_p g^2} -\frac{1}{2}\gamma_{gs} T_{sp} -\frac{1}{4}\left( \phi _R ^2 + \phi _p ^2 \right) & -\frac{1}{2} \frac{\kappa \gamma}{g^2}-\frac{1}{2}\frac{\kappa \sqrt{\log 2}}{\pi T_p g^2}-i\frac{g^2}{\kappa \Delta}  +\frac{1}{2}\left( \phi _R - \phi _p  \right) \\
-\frac{1}{2} \frac{\kappa \gamma}{g^2}-\frac{1}{2}\frac{\kappa \sqrt{\log 2}}{\pi T_p g^2}+i\frac{g^2}{\kappa \Delta}  +\frac{1}{2}\left( \phi _R - \phi _p  \right) & \frac{1}{2}\gamma_{gs} T_{sp} +\frac{1}{4}\left( \phi _R ^2 + \phi _p ^2 \right) \end{array} \right).
\end{align}
\end{widetext}
Defining the fidelity as
\begin{align}
F = \min \sqrt{\bra{\phi _{ideal}}\rho_a \ket{\phi_{ideal}}} \label{eq:fidelitydef}
\end{align}
where the ideal output state is given by Eq.(\ref{eq:ideal}).
\begin{align}
\ket{\phi _{ideal}} = \ket{0}_a \ket{1 }_p.
\end{align}
Then expanding the square root of Eq.(\ref{eq:fidelitydef}) and keeping the lowest order term in each correction, the fidelity is approximately
\begin{align}
F = \eta_{det} (&1 - \frac{\kappa \gamma}{2g^2} -\frac{\kappa \sqrt{\log 2}}{2\pi T_p g^2} -\frac{1}{4} \gamma _{gs} T_{sp}\nonumber \\
& - \frac{1}{8} (\phi_R^2 +  \phi_P^2) ), \label{eq:fidelity}
\end{align}
where $\eta_{det}$ is the efficiency of detecting the ideal state as discussed in Sect.\ref{sect:readout}. The fidelity is reduced by $\frac{\kappa \gamma}{2g^2}$ due to imperfect reflection of the photon, by $\frac{\kappa \sqrt{\log 2}}{2\pi T_p g^2}$ due to the finite bandwidth of the reflected photon, by
$\frac{1}{4} \gamma _{gs} T_{sp}$ due to dephasing while the atom is in the superposition state, and by $\frac{1}{8} (\phi_R ^2 + \phi_P^2 )$ due to
imperfect rotations when realizing the CPHASE gate.

For high fidelity we need high cooperativity $C = g^2/(\kappa \gamma) \gg 1$ which implies we need high quality
cavities. But the main limitation on the fidelity is the combination of needing
the factor $\gamma _{gs} T_{sp}$ to be small  while the factor $\kappa \sqrt{\log 2}/(2\pi T_p g^2)$ puts a lower limit on the pulse duration, such that the photon spectrum fits into the narrow bandwidth of the resonant feature.
The combination of these two factors will limit the overall fidelity, leading to one ideal pulse time, since $T_p$ is bound on both sides.
The last term due to imperfect rotations is actually quite small, if we make the cautious assumption that the area of the pulses is off by as much as 1\%,
then the fidelity is only reduced by 0.4\%, and likely the pulse areas can be made more accurate than that, so we can safely neglect this term.

\section{Implementation}
\label{sect:implement}
We need a single ion, strongly coupled to a cavity. Faraon et al. \cite{Faraon_2015,Faraon_2016} are
building photonic cavities which strongly couple to a number of rare-earth ions. 
In order to have a single ion coupled to the photonic cavity, the ion density can be lowered until
only a single ion couples to the cavity, but then the single ion may not be near the peak of the cavity mode,
and also may not have the right frequency.
There has also been work on using an ion beam to implant single REI into a pure crystal with Cerium ions implanted in YAG \cite{Kolesov_2014}
and Erbium ions implanted in YSO \cite{Bushev_2014}, which currently makes small spots of 1,000's
of REI, but could be scaled down to implanting a single REI.

In order to implement this protocol in a single rare earth ion, a long lived shelving state is needed, ideally a split ground state. This ground state splitting must be large enough such that
one of the states is far-detuned such that it can not interact with the photon and cavity while the other transition is in resonance.
Neodymium has a 9GHz separation in the presence of a 300mT magnetic field.
Such a large magnetic field is not necessary, but is routinely used.
We consider Neodymium because we have reliable data for it in a variety of crystals and coupling to a photonic crystal was already demonstrated,
but it could be that other ions will work just as well or better.

Nd:YVO$_4$ is an attractive implementation, since the Nd has a higher dipole moment in YVO$_4$, compared with YSO.
The energy diagram is shown Fig.\ref{fig:Ndlevel}.
High quality resonators which are capable of coupling to a single rare-earth ion, have recently been developed \cite{Faraon_2016}. The cavity has a mode volume of $V=(\lambda/n_{YVO})^3=0.064\mu m^3$ (where $\lambda=879.7nm$ is the Nd linewidth and $n_{YVO}=2.2$ is the refractive index of the YVO$_4$ crystal) and a quality factor of $Q=20,000$.
\begin{figure}
\includegraphics[width=6cm]{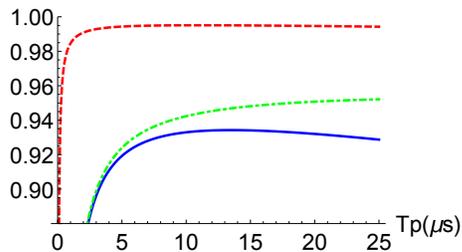}
\caption{(color online) Predicted fidelity for Nd:YVO$_4$ as a function of $T_p$ the time width of the pulse to be detected,
for the parameters listed in the text. The blue (solid) line is the fidelity for the experimentally demonstrated numbers listed in the text, the green (dot-dashed) line is for experimentally demonstrated numbers except with the lowest observed spin decoherence rate $\gamma_{gs} = 2\pi \cdot 0.34$kHz,
and the red (dashed) line is for a more optimistic theoretic quality factor of Q=300,000.} \label{fig:Ndfidelity}
\end{figure}
%
The electric field for a single photon in the cavity is given by:
\begin{align}
\mathcal{E} = \sqrt{\frac{\hbar \omega_c}{2 \epsilon _0 V}} \label{eq:field}
\end{align}
Where $\omega_c$ is the frequency of the cavity and $V$ is the cavity mode volume.  Then for Nd:YVO$_4$ we have $\mathcal{E} =446,229$V/m.
The HWHM linewidth of the cavity can be derived from the cavity frequency and quality factor.
\begin{align}
\kappa = \frac{1}{2} \frac{\omega _c}{Q}.
\end{align}
Then the cavity width is $\kappa = 2\pi \cdot 8.5$GHz.
The optical $T_2$ time for the Nd ion doped into YVO  was measured to be $27\mu$s with a 1.5T magnetic field \cite{Cone_2011}, which gives a decoherence rate of $\gamma = 2\pi \cdot5.9$kHz. With this field, the detuning can be as high as $\Delta = 2\pi \cdot30$GHz.
The transition in Nd:YVO$_4$ has a wavelength of 880nm, and according to \cite{Longdell_2009} has a dipole moment is  $\mu = 9.1 \cdot 10^{-32}$Cm.
Then the single photon Rabi frequency or the cavity-photon coupling is:
\begin{align}
g = \frac{\mu \mathcal{E}}{2\hbar},
\end{align}
such that $g=2\pi \cdot 30.6$MHz.
Then the single ion cooperativity $C = g^2/(\kappa \gamma) = 246$.
We can also calculate the Purcell factor using Eq.(\ref{eq:Purcell}), which is $F_P = 1520$.
By comparing the radiative decay rate for the transition and the total lifetime of the excited state we can find that the branching ratio from the excited state to the ground state is $\beta = 10.4\%$,
then the probability of emitting into the cavity is $p_{cav} = 0.9985$, thus the detection efficiency to detect a minimum of two photons, assuming the probability of detecting a single photon is $p_{det}=0.9$, given by Eq.(\ref{eq:detect}) is $\eta_{det} = 98.8\%$.

The Nd electron spin lifetime is quite long in Nd:YSO it was measured at $T_1 = 100$ms \cite{Goldner_2015}, giving ample time for the read-out process before
the excitation in $\ket{0}_a$ decays. 
The Nd electron spin coherence time is shorter at $T_2 = 471\mu$s at 5K \cite{Goldner_2015}, which gives
$\gamma_{gs} = 2\pi \cdot 0.34$kHz. At low temperatures, similar values should be possible for Nd:YVO$_4$. 
At the same time our pulses need to be long enough to fit into the limited bandwidth  of the resonant response from Eq.(\ref{eq:dzbc2}), $1/T_p \ll \delta \omega = 2\pi \cdot 1.3$MHz, so we have to make a compromise in pulse duration.
We plot the fidelity as a function of $T_p$ in Fig.\ref{fig:Ndfidelity}, here showing that pulses of length $T_p = 13\mu$s are ideal.
Then Eq.(\ref{eq:fidelity}) gives a maximum fidelity of $F=93.4$\%.
At sub-Kelvin temperatures the spin coherence time can be an order of magnitude higher $\gamma_{gs} = 2\pi \cdot 34$Hz, which improves the fidelity to $F=95.3$\% as shown in Fig.\ref{fig:Ndfidelity}. 

We also need to make sure the chance of false positives given by Eq.(\ref{eq:off}) is very low. For Nd:YVO$_4$ the detuning is large, the
width of the ion is small, but both transitions are equally allowed so $\tilde{g} = g$.
These combine to give a probability that is quite low at $p_{off} = 0.01\%$ and can safely be neglected.

There is no reason that cavities can not be improved to reach higher quality factors, in \cite{Faraon_2016} the theoretically possible quality factor is  $Q=300,000$ with the same mode volume $V=(\lambda/n_{YVO})^3$. Then $\kappa = 2\pi \cdot 565$MHz and the cavity-photon coupling is still $g=2\pi \cdot 30.6$MHz. Then the single ion cooperativity is C = 7392. The Purcell factor would be $F_P = 22,797$, giving a detection efficiency of $\eta = 99.1\%$.  Then with $\gamma_{gs} = 2\pi \cdot 34$Hz ideal pulse length is 11$\mu$s and the fidelity would be $F=99.5\%$, as shown in Fig.\ref{fig:Ndfidelity}.

\section{Conclusion}
We have demonstrated that a CPHASE gate between a single photon and a rare-earth doped ion coupled to a photonic cavity is possible in the bad cavity regime.
We have then shown that this gate can be used to make non-destructive measurements of a single photon. We suggested implementing a non-destructive photon measurement in Nd:YVO$_4$ and calculated the expected fidelity, concluding that high fidelities are within reach of current technology. A fidelity of 95.3\% is currently possible, and a theoretical maximum fidelity of 99.5\% could be achieved. Our results show that photonic crystal cavities coupled to individual rare-earth ions are a promising platform for implementing non-destructive photon detection in solid-state systems.

\section{Acknowledgments}
This work was supported by NSERC (Canada). AF and TZ acknowledge support from National Science Foundation CAREER award 1454607. We thank Dr. John Bartolomew for useful discussion.

\bibliographystyle{apsrev4-1}
\bibliography{Transducer_bib}

\begin{thebibliography}{34}%
\makeatletter
\providecommand \@ifxundefined [1]{%
 \@ifx{#1\undefined}
}%
\providecommand \@ifnum [1]{%
 \ifnum #1\expandafter \@firstoftwo
 \else \expandafter \@secondoftwo
 \fi
}%
\providecommand \@ifx [1]{%
 \ifx #1\expandafter \@firstoftwo
 \else \expandafter \@secondoftwo
 \fi
}%
\providecommand \natexlab [1]{#1}%
\providecommand \enquote  [1]{``#1''}%
\providecommand \bibnamefont  [1]{#1}%
\providecommand \bibfnamefont [1]{#1}%
\providecommand \citenamefont [1]{#1}%
\providecommand \href@noop [0]{\@secondoftwo}%
\providecommand \href [0]{\begingroup \@sanitize@url \@href}%
\providecommand \@href[1]{\@@startlink{#1}\@@href}%
\providecommand \@@href[1]{\endgroup#1\@@endlink}%
\providecommand \@sanitize@url [0]{\catcode `\\12\catcode `\$12\catcode
  `\&12\catcode `\#12\catcode `\^12\catcode `\_12\catcode `\%12\relax}%
\providecommand \@@startlink[1]{}%
\providecommand \@@endlink[0]{}%
\providecommand \url  [0]{\begingroup\@sanitize@url \@url }%
\providecommand \@url [1]{\endgroup\@href {#1}{\urlprefix }}%
\providecommand \urlprefix  [0]{URL }%
\providecommand \Eprint [0]{\href }%
\providecommand \doibase [0]{http://dx.doi.org/}%
\providecommand \selectlanguage [0]{\@gobble}%
\providecommand \bibinfo  [0]{\@secondoftwo}%
\providecommand \bibfield  [0]{\@secondoftwo}%
\providecommand \translation [1]{[#1]}%
\providecommand \BibitemOpen [0]{}%
\providecommand \bibitemStop [0]{}%
\providecommand \bibitemNoStop [0]{.\EOS\space}%
\providecommand \EOS [0]{\spacefactor3000\relax}%
\providecommand \BibitemShut  [1]{\csname bibitem#1\endcsname}%
\let\auto@bib@innerbib\@empty
\bibitem [{\citenamefont {Sangouard}\ \emph {et~al.}(2011)\citenamefont
  {Sangouard}, \citenamefont {Simon}, \citenamefont {De~Riedmatten},\ and\
  \citenamefont {Gisin}}]{Repreview}%
  \BibitemOpen
  \bibfield  {author} {\bibinfo {author} {\bibfnamefont {N.}~\bibnamefont
  {Sangouard}}, \bibinfo {author} {\bibfnamefont {C.}~\bibnamefont {Simon}},
  \bibinfo {author} {\bibfnamefont {H.}~\bibnamefont {De~Riedmatten}}, \ and\
  \bibinfo {author} {\bibfnamefont {N.}~\bibnamefont {Gisin}},\ }\href@noop {}
  {\bibfield  {journal} {\bibinfo  {journal} {Rev. Mod. Phys.}\ }\textbf
  {\bibinfo {volume} {83}},\ \bibinfo {pages} {33} (\bibinfo {year}
  {2011})}\BibitemShut {NoStop}%
\bibitem [{\citenamefont {Boone}\ \emph {et~al.}(2015)\citenamefont {Boone},
  \citenamefont {Bourgoin}, \citenamefont {Meyer-Scott}, \citenamefont
  {Heshami}, \citenamefont {Jennewein},\ and\ \citenamefont
  {Simon}}]{satellite}%
  \BibitemOpen
  \bibfield  {author} {\bibinfo {author} {\bibfnamefont {K.}~\bibnamefont
  {Boone}}, \bibinfo {author} {\bibfnamefont {J.-P.}\ \bibnamefont {Bourgoin}},
  \bibinfo {author} {\bibfnamefont {E.}~\bibnamefont {Meyer-Scott}}, \bibinfo
  {author} {\bibfnamefont {K.}~\bibnamefont {Heshami}}, \bibinfo {author}
  {\bibfnamefont {T.}~\bibnamefont {Jennewein}}, \ and\ \bibinfo {author}
  {\bibfnamefont {C.}~\bibnamefont {Simon}},\ }\href {\doibase
  10.1103/PhysRevA.91.052325} {\bibfield  {journal} {\bibinfo  {journal} {Phys.
  Rev. A}\ }\textbf {\bibinfo {volume} {91}},\ \bibinfo {pages} {052325}
  (\bibinfo {year} {2015})}\BibitemShut {NoStop}%
\bibitem [{\citenamefont {Knill}\ \emph {et~al.}(2000)\citenamefont {Knill},
  \citenamefont {Laflamme},\ and\ \citenamefont {Milburn}}]{PQCtheory}%
  \BibitemOpen
  \bibfield  {author} {\bibinfo {author} {\bibfnamefont {E.}~\bibnamefont
  {Knill}}, \bibinfo {author} {\bibfnamefont {R.}~\bibnamefont {Laflamme}}, \
  and\ \bibinfo {author} {\bibfnamefont {G.~J.}\ \bibnamefont {Milburn}},\
  }\href {http://www.nature.com/nature/journal/v409/n6816/full/409046a0.html}
  {\bibfield  {journal} {\bibinfo  {journal} {Nature}\ }\textbf {\bibinfo
  {volume} {409}},\ \bibinfo {pages} {46} (\bibinfo {year} {2000})}\BibitemShut
  {NoStop}%
\bibitem [{\citenamefont {Humphreys}\ \emph {et~al.}(2013)\citenamefont
  {Humphreys}, \citenamefont {Metcalf}, \citenamefont {Spring}, \citenamefont
  {Moore}, \citenamefont {Jin}, \citenamefont {Barbieri}, \citenamefont
  {Kolthammer},\ and\ \citenamefont {Walmsley}}]{PQCexp}%
  \BibitemOpen
  \bibfield  {author} {\bibinfo {author} {\bibfnamefont {P.~C.}\ \bibnamefont
  {Humphreys}}, \bibinfo {author} {\bibfnamefont {B.~J.}\ \bibnamefont
  {Metcalf}}, \bibinfo {author} {\bibfnamefont {J.~B.}\ \bibnamefont {Spring}},
  \bibinfo {author} {\bibfnamefont {M.}~\bibnamefont {Moore}}, \bibinfo
  {author} {\bibfnamefont {X.-M.}\ \bibnamefont {Jin}}, \bibinfo {author}
  {\bibfnamefont {M.}~\bibnamefont {Barbieri}}, \bibinfo {author}
  {\bibfnamefont {W.~S.}\ \bibnamefont {Kolthammer}}, \ and\ \bibinfo {author}
  {\bibfnamefont {I.~A.}\ \bibnamefont {Walmsley}},\ }\href {\doibase
  10.1103/PhysRevLett.111.150501} {\bibfield  {journal} {\bibinfo  {journal}
  {Phys. Rev. Lett.}\ }\textbf {\bibinfo {volume} {111}},\ \bibinfo {pages}
  {150501} (\bibinfo {year} {2013})}\BibitemShut {NoStop}%
\bibitem [{\citenamefont {O'Brien}\ \emph {et~al.}(2003)\citenamefont
  {O'Brien}, \citenamefont {Pryde}, \citenamefont {White}, \citenamefont
  {Ralph},\ and\ \citenamefont {Branning}}]{OBrien_2003}%
  \BibitemOpen
  \bibfield  {author} {\bibinfo {author} {\bibfnamefont {J.}~\bibnamefont
  {O'Brien}}, \bibinfo {author} {\bibfnamefont {G.}~\bibnamefont {Pryde}},
  \bibinfo {author} {\bibfnamefont {A.}~\bibnamefont {White}}, \bibinfo
  {author} {\bibfnamefont {T.}~\bibnamefont {Ralph}}, \ and\ \bibinfo {author}
  {\bibfnamefont {D.}~\bibnamefont {Branning}},\ }\href@noop {} {\bibfield
  {journal} {\bibinfo  {journal} {Nature}\ }\textbf {\bibinfo {volume} {426}},\
  \bibinfo {pages} {264} (\bibinfo {year} {2003})}\BibitemShut {NoStop}%
\bibitem [{\citenamefont {Turchette}\ \emph {et~al.}(1998)\citenamefont
  {Turchette}, \citenamefont {Georgiades}, \citenamefont {Hood}, \citenamefont
  {Kimble},\ and\ \citenamefont {Parkins}}]{CPHASEexp}%
  \BibitemOpen
  \bibfield  {author} {\bibinfo {author} {\bibfnamefont {Q.~A.}\ \bibnamefont
  {Turchette}}, \bibinfo {author} {\bibfnamefont {N.~P.}\ \bibnamefont
  {Georgiades}}, \bibinfo {author} {\bibfnamefont {C.~J.}\ \bibnamefont
  {Hood}}, \bibinfo {author} {\bibfnamefont {H.~J.}\ \bibnamefont {Kimble}}, \
  and\ \bibinfo {author} {\bibfnamefont {A.~S.}\ \bibnamefont {Parkins}},\
  }\href {\doibase 10.1103/PhysRevA.58.4056} {\bibfield  {journal} {\bibinfo
  {journal} {Phys. Rev. A}\ }\textbf {\bibinfo {volume} {58}},\ \bibinfo
  {pages} {4056} (\bibinfo {year} {1998})}\BibitemShut {NoStop}%
\bibitem [{\citenamefont {Duan}\ and\ \citenamefont
  {Kimble}(2004)}]{Duan_2004}%
  \BibitemOpen
  \bibfield  {author} {\bibinfo {author} {\bibfnamefont {L.-M.}\ \bibnamefont
  {Duan}}\ and\ \bibinfo {author} {\bibfnamefont {H.~J.}\ \bibnamefont
  {Kimble}},\ }\href {http://link.aps.org/doi/10.1103/PhysRevLett.92.127902}
  {\bibfield  {journal} {\bibinfo  {journal} {Phys. Rev. Lett.}\ }\textbf
  {\bibinfo {volume} {92}},\ \bibinfo {pages} {127902} (\bibinfo {year}
  {2004})}\BibitemShut {NoStop}%
\bibitem [{\citenamefont {Reiserer}\ \emph {et~al.}(2013)\citenamefont
  {Reiserer}, \citenamefont {Ritter},\ and\ \citenamefont
  {Rempe}}]{Rempe_2013}%
  \BibitemOpen
  \bibfield  {author} {\bibinfo {author} {\bibfnamefont {A.}~\bibnamefont
  {Reiserer}}, \bibinfo {author} {\bibfnamefont {S.}~\bibnamefont {Ritter}}, \
  and\ \bibinfo {author} {\bibfnamefont {G.}~\bibnamefont {Rempe}},\
  }\href@noop {} {\bibfield  {journal} {\bibinfo  {journal} {Science}\ }\textbf
  {\bibinfo {volume} {342}},\ \bibinfo {pages} {1349} (\bibinfo {year}
  {2013})}\BibitemShut {NoStop}%
\bibitem [{\citenamefont {Reiserer}\ \emph {et~al.}(2014)\citenamefont
  {Reiserer}, \citenamefont {Kalb}, \citenamefont {Rempe},\ and\ \citenamefont
  {Ritter}}]{Rempe_2014}%
  \BibitemOpen
  \bibfield  {author} {\bibinfo {author} {\bibfnamefont {A.}~\bibnamefont
  {Reiserer}}, \bibinfo {author} {\bibfnamefont {N.}~\bibnamefont {Kalb}},
  \bibinfo {author} {\bibfnamefont {G.}~\bibnamefont {Rempe}}, \ and\ \bibinfo
  {author} {\bibfnamefont {S.}~\bibnamefont {Ritter}},\ }\href@noop {}
  {\bibfield  {journal} {\bibinfo  {journal} {Nature}\ }\textbf {\bibinfo
  {volume} {508}},\ \bibinfo {pages} {237} (\bibinfo {year}
  {2014})}\BibitemShut {NoStop}%
\bibitem [{\citenamefont {Kalb}\ \emph {et~al.}(2015)\citenamefont {Kalb},
  \citenamefont {Reiserer}, \citenamefont {Ritter},\ and\ \citenamefont
  {Rempe}}]{Rempe_2015}%
  \BibitemOpen
  \bibfield  {author} {\bibinfo {author} {\bibfnamefont {N.}~\bibnamefont
  {Kalb}}, \bibinfo {author} {\bibfnamefont {A.}~\bibnamefont {Reiserer}},
  \bibinfo {author} {\bibfnamefont {S.}~\bibnamefont {Ritter}}, \ and\ \bibinfo
  {author} {\bibfnamefont {G.}~\bibnamefont {Rempe}},\ }\href@noop {}
  {\bibfield  {journal} {\bibinfo  {journal} {Phys. Rev. Lett.}\ }\textbf
  {\bibinfo {volume} {114}},\ \bibinfo {pages} {220501} (\bibinfo {year}
  {2015})}\BibitemShut {NoStop}%
\bibitem [{\citenamefont {Hacker}\ \emph {et~al.}(2016)\citenamefont {Hacker},
  \citenamefont {Welte}, \citenamefont {Rempe},\ and\ \citenamefont
  {Ritter}}]{Rempe_2016}%
  \BibitemOpen
  \bibfield  {author} {\bibinfo {author} {\bibfnamefont {B.}~\bibnamefont
  {Hacker}}, \bibinfo {author} {\bibfnamefont {S.}~\bibnamefont {Welte}},
  \bibinfo {author} {\bibfnamefont {G.}~\bibnamefont {Rempe}}, \ and\ \bibinfo
  {author} {\bibfnamefont {S.}~\bibnamefont {Ritter}},\ }\href@noop {}
  {\bibfield  {journal} {\bibinfo  {journal} {arXiv:}\ }\textbf {\bibinfo
  {volume} {1605.05261}} (\bibinfo {year} {2016})}\BibitemShut {NoStop}%
\bibitem [{\citenamefont {Lauritzen}\ \emph {et~al.}(2010)\citenamefont
  {Lauritzen}, \citenamefont {Min\'a$\check{r}$}, \citenamefont
  {de~Riedmatten}, \citenamefont {Afzelius}, \citenamefont {Sangouard},
  \citenamefont {Simon},\ and\ \citenamefont {Gisin}}]{GisinQM}%
  \BibitemOpen
  \bibfield  {author} {\bibinfo {author} {\bibfnamefont {B.}~\bibnamefont
  {Lauritzen}}, \bibinfo {author} {\bibfnamefont {J.}~\bibnamefont
  {Min\'a$\check{r}$}}, \bibinfo {author} {\bibfnamefont {H.}~\bibnamefont
  {de~Riedmatten}}, \bibinfo {author} {\bibfnamefont {M.}~\bibnamefont
  {Afzelius}}, \bibinfo {author} {\bibfnamefont {N.}~\bibnamefont {Sangouard}},
  \bibinfo {author} {\bibfnamefont {C.}~\bibnamefont {Simon}}, \ and\ \bibinfo
  {author} {\bibfnamefont {N.}~\bibnamefont {Gisin}},\ }\href {\doibase
  10.1103/PhysRevLett.104.080502} {\bibfield  {journal} {\bibinfo  {journal}
  {Phys. Rev. Lett.}\ }\textbf {\bibinfo {volume} {104}},\ \bibinfo {pages}
  {080502} (\bibinfo {year} {2010})}\BibitemShut {NoStop}%
\bibitem [{\citenamefont {Afzelius}\ \emph
  {et~al.}(2010{\natexlab{a}})\citenamefont {Afzelius}, \citenamefont {Usmani},
  \citenamefont {Amari}, \citenamefont {Lauritzen}, \citenamefont {Walther},
  \citenamefont {Simon}, \citenamefont {Sangouard}, \citenamefont
  {Min\'a$\check{r}$}, \citenamefont {de~Riedmatten}, \citenamefont {Gisin},\
  and\ \citenamefont {Kr\"oll}}]{KrollQM}%
  \BibitemOpen
  \bibfield  {author} {\bibinfo {author} {\bibfnamefont {M.}~\bibnamefont
  {Afzelius}}, \bibinfo {author} {\bibfnamefont {I.}~\bibnamefont {Usmani}},
  \bibinfo {author} {\bibfnamefont {A.}~\bibnamefont {Amari}}, \bibinfo
  {author} {\bibfnamefont {B.}~\bibnamefont {Lauritzen}}, \bibinfo {author}
  {\bibfnamefont {A.}~\bibnamefont {Walther}}, \bibinfo {author} {\bibfnamefont
  {C.}~\bibnamefont {Simon}}, \bibinfo {author} {\bibfnamefont
  {N.}~\bibnamefont {Sangouard}}, \bibinfo {author} {\bibfnamefont
  {J.}~\bibnamefont {Min\'a$\check{r}$}}, \bibinfo {author} {\bibfnamefont
  {H.}~\bibnamefont {de~Riedmatten}}, \bibinfo {author} {\bibfnamefont
  {N.}~\bibnamefont {Gisin}}, \ and\ \bibinfo {author} {\bibfnamefont
  {S.}~\bibnamefont {Kr\"oll}},\ }\href {\doibase
  10.1103/PhysRevLett.104.040503} {\bibfield  {journal} {\bibinfo  {journal}
  {Phys. Rev. Lett.}\ }\textbf {\bibinfo {volume} {104}},\ \bibinfo {pages}
  {040503} (\bibinfo {year} {2010}{\natexlab{a}})}\BibitemShut {NoStop}%
\bibitem [{\citenamefont {Hedges}\ \emph {et~al.}(2010)\citenamefont {Hedges},
  \citenamefont {Longdell}, \citenamefont {Li},\ and\ \citenamefont
  {Sellars}}]{LongdellQM}%
  \BibitemOpen
  \bibfield  {author} {\bibinfo {author} {\bibfnamefont {M.~P.}\ \bibnamefont
  {Hedges}}, \bibinfo {author} {\bibfnamefont {J.~J.}\ \bibnamefont
  {Longdell}}, \bibinfo {author} {\bibfnamefont {Y.}~\bibnamefont {Li}}, \ and\
  \bibinfo {author} {\bibfnamefont {M.~J.}\ \bibnamefont {Sellars}},\ }\href
  {http://www.nature.com/nature/journal/v465/n7301/suppinfo/nature09081_S1.html}
  {\bibfield  {journal} {\bibinfo  {journal} {Nature}\ }\textbf {\bibinfo
  {volume} {465}},\ \bibinfo {pages} {1052} (\bibinfo {year}
  {2010})}\BibitemShut {NoStop}%
\bibitem [{\citenamefont {Wesenberg}\ \emph {et~al.}(2007)\citenamefont
  {Wesenberg}, \citenamefont {M\o{}lmer}, \citenamefont {Rippe},\ and\
  \citenamefont {Kr\"oll}}]{Kroll_2007}%
  \BibitemOpen
  \bibfield  {author} {\bibinfo {author} {\bibfnamefont {J.~H.}\ \bibnamefont
  {Wesenberg}}, \bibinfo {author} {\bibfnamefont {K.}~\bibnamefont
  {M\o{}lmer}}, \bibinfo {author} {\bibfnamefont {L.}~\bibnamefont {Rippe}}, \
  and\ \bibinfo {author} {\bibfnamefont {S.}~\bibnamefont {Kr\"oll}},\ }\href
  {\doibase 10.1103/PhysRevA.75.012304} {\bibfield  {journal} {\bibinfo
  {journal} {Phys. Rev. A}\ }\textbf {\bibinfo {volume} {75}},\ \bibinfo
  {pages} {012304} (\bibinfo {year} {2007})}\BibitemShut {NoStop}%
\bibitem [{\citenamefont {Sinclair}\ \emph {et~al.}(2015)\citenamefont
  {Sinclair}, \citenamefont {Heshami}, \citenamefont {Deshmukh}, \citenamefont
  {Oblak}, \citenamefont {Simon},\ and\ \citenamefont
  {Tittel}}]{Sinclair_2015}%
  \BibitemOpen
  \bibfield  {author} {\bibinfo {author} {\bibfnamefont {N.}~\bibnamefont
  {Sinclair}}, \bibinfo {author} {\bibfnamefont {K.}~\bibnamefont {Heshami}},
  \bibinfo {author} {\bibfnamefont {C.}~\bibnamefont {Deshmukh}}, \bibinfo
  {author} {\bibfnamefont {D.}~\bibnamefont {Oblak}}, \bibinfo {author}
  {\bibfnamefont {C.}~\bibnamefont {Simon}}, \ and\ \bibinfo {author}
  {\bibfnamefont {W.}~\bibnamefont {Tittel}},\ }\href@noop {} {\bibfield
  {journal} {\bibinfo  {journal} {arXiv:}\ }\textbf {\bibinfo {volume}
  {1510.01164}} (\bibinfo {year} {2015})}\BibitemShut {NoStop}%
\bibitem [{\citenamefont {Kolesov}\ \emph {et~al.}(2012)\citenamefont
  {Kolesov}, \citenamefont {Xia}, \citenamefont {Reuter}, \citenamefont
  {St\"ohr}, \citenamefont {Zappe}, \citenamefont {Meijer}, \citenamefont
  {Hemmer},\ and\ \citenamefont {Wratchtrup}}]{Kolesov_2012}%
  \BibitemOpen
  \bibfield  {author} {\bibinfo {author} {\bibfnamefont {R.}~\bibnamefont
  {Kolesov}}, \bibinfo {author} {\bibfnamefont {K.}~\bibnamefont {Xia}},
  \bibinfo {author} {\bibfnamefont {R.}~\bibnamefont {Reuter}}, \bibinfo
  {author} {\bibfnamefont {R.}~\bibnamefont {St\"ohr}}, \bibinfo {author}
  {\bibfnamefont {A.}~\bibnamefont {Zappe}}, \bibinfo {author} {\bibfnamefont
  {J.}~\bibnamefont {Meijer}}, \bibinfo {author} {\bibfnamefont
  {P.}~\bibnamefont {Hemmer}}, \ and\ \bibinfo {author} {\bibfnamefont
  {J.}~\bibnamefont {Wratchtrup}},\ }\href@noop {} {\bibfield  {journal}
  {\bibinfo  {journal} {Nat. Comm.}\ }\textbf {\bibinfo {volume} {3}},\
  \bibinfo {pages} {1029} (\bibinfo {year} {2012})}\BibitemShut {NoStop}%
\bibitem [{\citenamefont {Yin}\ \emph {et~al.}(2013)\citenamefont {Yin},
  \citenamefont {Rancie}, \citenamefont {de~Boo}, \citenamefont {Stavrias},
  \citenamefont {McCallum}, \citenamefont {Sellars},\ and\ \citenamefont
  {Rogge}}]{Rogge_2013}%
  \BibitemOpen
  \bibfield  {author} {\bibinfo {author} {\bibfnamefont {C.}~\bibnamefont
  {Yin}}, \bibinfo {author} {\bibfnamefont {M.}~\bibnamefont {Rancie}},
  \bibinfo {author} {\bibfnamefont {G.~G.}\ \bibnamefont {de~Boo}}, \bibinfo
  {author} {\bibfnamefont {N.}~\bibnamefont {Stavrias}}, \bibinfo {author}
  {\bibfnamefont {J.~C.}\ \bibnamefont {McCallum}}, \bibinfo {author}
  {\bibfnamefont {M.~J.}\ \bibnamefont {Sellars}}, \ and\ \bibinfo {author}
  {\bibfnamefont {S.}~\bibnamefont {Rogge}},\ }\href@noop {} {\bibfield
  {journal} {\bibinfo  {journal} {Nature}\ }\textbf {\bibinfo {volume} {497}},\
  \bibinfo {pages} {91} (\bibinfo {year} {2013})}\BibitemShut {NoStop}%
\bibitem [{\citenamefont {Ahlefeldt}\ \emph {et~al.}(2013)\citenamefont
  {Ahlefeldt}, \citenamefont {McAuslan}, \citenamefont {Longdell},
  \citenamefont {Manson},\ and\ \citenamefont {Sellars}}]{Sellars_2013}%
  \BibitemOpen
  \bibfield  {author} {\bibinfo {author} {\bibfnamefont {R.~L.}\ \bibnamefont
  {Ahlefeldt}}, \bibinfo {author} {\bibfnamefont {D.~L.}\ \bibnamefont
  {McAuslan}}, \bibinfo {author} {\bibfnamefont {J.~J.}\ \bibnamefont
  {Longdell}}, \bibinfo {author} {\bibfnamefont {N.~B.}\ \bibnamefont
  {Manson}}, \ and\ \bibinfo {author} {\bibfnamefont {M.~J.}\ \bibnamefont
  {Sellars}},\ }\href {http://link.aps.org/doi/10.1103/PhysRevLett.111.240501}
  {\bibfield  {journal} {\bibinfo  {journal} {Phys. Rev. Lett.}\ }\textbf
  {\bibinfo {volume} {111}},\ \bibinfo {pages} {240501} (\bibinfo {year}
  {2013})}\BibitemShut {NoStop}%
\bibitem [{\citenamefont {Utikal}\ \emph {et~al.}(2014)\citenamefont {Utikal},
  \citenamefont {Eichhammer}, \citenamefont {Petersen}, \citenamefont {Renn},
  \citenamefont {G\"otzinger},\ and\ \citenamefont
  {Sandoghdar}}]{Sandoghdar_2014}%
  \BibitemOpen
  \bibfield  {author} {\bibinfo {author} {\bibfnamefont {T.}~\bibnamefont
  {Utikal}}, \bibinfo {author} {\bibfnamefont {E.}~\bibnamefont {Eichhammer}},
  \bibinfo {author} {\bibfnamefont {L.}~\bibnamefont {Petersen}}, \bibinfo
  {author} {\bibfnamefont {A.}~\bibnamefont {Renn}}, \bibinfo {author}
  {\bibfnamefont {S.}~\bibnamefont {G\"otzinger}}, \ and\ \bibinfo {author}
  {\bibfnamefont {V.}~\bibnamefont {Sandoghdar}},\ }\href@noop {} {\bibfield
  {journal} {\bibinfo  {journal} {Nat. Comm.}\ }\textbf {\bibinfo {volume}
  {5}},\ \bibinfo {pages} {3627} (\bibinfo {year} {2014})}\BibitemShut
  {NoStop}%
\bibitem [{\citenamefont {Eichhammer}\ \emph {et~al.}(2015)\citenamefont
  {Eichhammer}, \citenamefont {Utikal}, \citenamefont {G\"otzinger},\ and\
  \citenamefont {Sandoghdar}}]{Sandoghdar_2015}%
  \BibitemOpen
  \bibfield  {author} {\bibinfo {author} {\bibfnamefont {E.}~\bibnamefont
  {Eichhammer}}, \bibinfo {author} {\bibfnamefont {T.}~\bibnamefont {Utikal}},
  \bibinfo {author} {\bibfnamefont {S.}~\bibnamefont {G\"otzinger}}, \ and\
  \bibinfo {author} {\bibfnamefont {V.}~\bibnamefont {Sandoghdar}},\ }\href
  {http://stacks.iop.org/1367-2630/17/i=8/a=083018} {\bibfield  {journal}
  {\bibinfo  {journal} {N. Jour. Phys.}\ }\textbf {\bibinfo {volume} {17}},\
  \bibinfo {pages} {083018} (\bibinfo {year} {2015})}\BibitemShut {NoStop}%
\bibitem [{\citenamefont {Kolesov}\ \emph {et~al.}(2013)\citenamefont
  {Kolesov}, \citenamefont {Xia}, \citenamefont {Reuter}, \citenamefont
  {Jamali}, \citenamefont {St\"ohr}, \citenamefont {Inal}, \citenamefont
  {Siyushev},\ and\ \citenamefont {Wrachtrup}}]{Kolesov_2013}%
  \BibitemOpen
  \bibfield  {author} {\bibinfo {author} {\bibfnamefont {R.}~\bibnamefont
  {Kolesov}}, \bibinfo {author} {\bibfnamefont {K.}~\bibnamefont {Xia}},
  \bibinfo {author} {\bibfnamefont {R.}~\bibnamefont {Reuter}}, \bibinfo
  {author} {\bibfnamefont {M.}~\bibnamefont {Jamali}}, \bibinfo {author}
  {\bibfnamefont {R.}~\bibnamefont {St\"ohr}}, \bibinfo {author} {\bibfnamefont
  {T.}~\bibnamefont {Inal}}, \bibinfo {author} {\bibfnamefont {P.}~\bibnamefont
  {Siyushev}}, \ and\ \bibinfo {author} {\bibfnamefont {J.}~\bibnamefont
  {Wrachtrup}},\ }\href
  {http://link.aps.org/doi/10.1103/PhysRevLett.111.120502} {\bibfield
  {journal} {\bibinfo  {journal} {Phys. Rev. Lett.}\ }\textbf {\bibinfo
  {volume} {111}},\ \bibinfo {pages} {120502} (\bibinfo {year}
  {2013})}\BibitemShut {NoStop}%
\bibitem [{\citenamefont {Zhong}\ \emph {et~al.}(2015)\citenamefont {Zhong},
  \citenamefont {Kindem}, \citenamefont {Miyazono},\ and\ \citenamefont
  {Faraon}}]{Faraon_2015}%
  \BibitemOpen
  \bibfield  {author} {\bibinfo {author} {\bibfnamefont {T.}~\bibnamefont
  {Zhong}}, \bibinfo {author} {\bibfnamefont {J.~M.}\ \bibnamefont {Kindem}},
  \bibinfo {author} {\bibfnamefont {E.}~\bibnamefont {Miyazono}}, \ and\
  \bibinfo {author} {\bibfnamefont {A.}~\bibnamefont {Faraon}},\ }\href@noop {}
  {\bibfield  {journal} {\bibinfo  {journal} {Nat. Comm.}\ }\textbf {\bibinfo
  {volume} {6}},\ \bibinfo {pages} {8206} (\bibinfo {year} {2015})}\BibitemShut
  {NoStop}%
\bibitem [{\citenamefont {Zhong}\ \emph {et~al.}(2016)\citenamefont {Zhong},
  \citenamefont {Rochman}, \citenamefont {Kindem}, \citenamefont {Miyazono},\
  and\ \citenamefont {Faraon}}]{Faraon_2016}%
  \BibitemOpen
  \bibfield  {author} {\bibinfo {author} {\bibfnamefont {T.}~\bibnamefont
  {Zhong}}, \bibinfo {author} {\bibfnamefont {J.}~\bibnamefont {Rochman}},
  \bibinfo {author} {\bibfnamefont {J.}~\bibnamefont {Kindem}}, \bibinfo
  {author} {\bibfnamefont {E.}~\bibnamefont {Miyazono}}, \ and\ \bibinfo
  {author} {\bibfnamefont {A.}~\bibnamefont {Faraon}},\ }\href@noop {}
  {\bibfield  {journal} {\bibinfo  {journal} {Opt. Exp.}\ }\textbf {\bibinfo
  {volume} {24}},\ \bibinfo {pages} {536} (\bibinfo {year} {2016})}\BibitemShut
  {NoStop}%
\bibitem [{\citenamefont {McAuslan}\ \emph {et~al.}(2011)\citenamefont
  {McAuslan}, \citenamefont {Korystov},\ and\ \citenamefont
  {Longdell}}]{Longdell_2009}%
  \BibitemOpen
  \bibfield  {author} {\bibinfo {author} {\bibfnamefont {D.~L.}\ \bibnamefont
  {McAuslan}}, \bibinfo {author} {\bibfnamefont {D.}~\bibnamefont {Korystov}},
  \ and\ \bibinfo {author} {\bibfnamefont {J.~J.}\ \bibnamefont {Longdell}},\
  }\href {http://link.aps.org/doi/10.1103/PhysRevA.83.063847} {\bibfield
  {journal} {\bibinfo  {journal} {Phys. Rev. A}\ }\textbf {\bibinfo {volume}
  {83}},\ \bibinfo {pages} {063847} (\bibinfo {year} {2011})}\BibitemShut
  {NoStop}%
\bibitem [{\citenamefont {Anisimov}\ and\ \citenamefont
  {Kocharovskaya}(2008)}]{Petr}%
  \BibitemOpen
  \bibfield  {author} {\bibinfo {author} {\bibfnamefont {P.}~\bibnamefont
  {Anisimov}}\ and\ \bibinfo {author} {\bibfnamefont {O.}~\bibnamefont
  {Kocharovskaya}},\ }\href@noop {} {\bibfield  {journal} {\bibinfo  {journal}
  {J. Mod. Opt.}\ }\textbf {\bibinfo {volume} {55}},\ \bibinfo {pages} {3159}
  (\bibinfo {year} {2008})}\BibitemShut {NoStop}%
\bibitem [{\citenamefont {O’Brien}\ \emph {et~al.}(2011)\citenamefont
  {O’Brien}, \citenamefont {Anisimov}, \citenamefont {Rostovtsev},\ and\
  \citenamefont {Kocharovskaya}}]{OBrienRb}%
  \BibitemOpen
  \bibfield  {author} {\bibinfo {author} {\bibfnamefont {C.}~\bibnamefont
  {O’Brien}}, \bibinfo {author} {\bibfnamefont {P.~M.}\ \bibnamefont
  {Anisimov}}, \bibinfo {author} {\bibfnamefont {Y.}~\bibnamefont
  {Rostovtsev}}, \ and\ \bibinfo {author} {\bibfnamefont {O.}~\bibnamefont
  {Kocharovskaya}},\ }\href@noop {} {\bibfield  {journal} {\bibinfo  {journal}
  {Physical Review A}\ }\textbf {\bibinfo {volume} {84}},\ \bibinfo {pages}
  {063835} (\bibinfo {year} {2011})}\BibitemShut {NoStop}%
\bibitem [{\citenamefont {Rippe}\ \emph {et~al.}(2005)\citenamefont {Rippe},
  \citenamefont {Nilsson}, \citenamefont {Kr\"oll}, \citenamefont {Klieber},\
  and\ \citenamefont {Suter}}]{transfer}%
  \BibitemOpen
  \bibfield  {author} {\bibinfo {author} {\bibfnamefont {L.}~\bibnamefont
  {Rippe}}, \bibinfo {author} {\bibfnamefont {M.}~\bibnamefont {Nilsson}},
  \bibinfo {author} {\bibfnamefont {S.}~\bibnamefont {Kr\"oll}}, \bibinfo
  {author} {\bibfnamefont {R.}~\bibnamefont {Klieber}}, \ and\ \bibinfo
  {author} {\bibfnamefont {D.}~\bibnamefont {Suter}},\ }\href {\doibase
  10.1103/PhysRevA.71.062328} {\bibfield  {journal} {\bibinfo  {journal} {Phys.
  Rev. A}\ }\textbf {\bibinfo {volume} {71}},\ \bibinfo {pages} {062328}
  (\bibinfo {year} {2005})}\BibitemShut {NoStop}%
\bibitem [{\citenamefont {Afzelius}\ \emph
  {et~al.}(2010{\natexlab{b}})\citenamefont {Afzelius}, \citenamefont {Staudt},
  \citenamefont {de~Riedmatten}, \citenamefont {Gisin}, \citenamefont
  {Guillot-Noël}, \citenamefont {Goldner}, \citenamefont {Marino},
  \citenamefont {Porcher}, \citenamefont {Cavalli},\ and\ \citenamefont
  {Bettinelli}}]{NDYVO}%
  \BibitemOpen
  \bibfield  {author} {\bibinfo {author} {\bibfnamefont {M.}~\bibnamefont
  {Afzelius}}, \bibinfo {author} {\bibfnamefont {M.~U.}\ \bibnamefont
  {Staudt}}, \bibinfo {author} {\bibfnamefont {H.}~\bibnamefont
  {de~Riedmatten}}, \bibinfo {author} {\bibfnamefont {N.}~\bibnamefont
  {Gisin}}, \bibinfo {author} {\bibfnamefont {O.}~\bibnamefont
  {Guillot-Noël}}, \bibinfo {author} {\bibfnamefont {P.}~\bibnamefont
  {Goldner}}, \bibinfo {author} {\bibfnamefont {R.}~\bibnamefont {Marino}},
  \bibinfo {author} {\bibfnamefont {P.}~\bibnamefont {Porcher}}, \bibinfo
  {author} {\bibfnamefont {E.}~\bibnamefont {Cavalli}}, \ and\ \bibinfo
  {author} {\bibfnamefont {M.}~\bibnamefont {Bettinelli}},\ }\href {\doibase
  http://dx.doi.org/10.1016/j.jlumin.2009.12.026} {\bibfield  {journal}
  {\bibinfo  {journal} {Journal of Luminescence}\ }\textbf {\bibinfo {volume}
  {130}},\ \bibinfo {pages} {1566 } (\bibinfo {year} {2010}{\natexlab{b}})},\
  \bibinfo {note} {special Issue based on the Proceedings of the Tenth
  International Meeting on Hole Burning, Single Molecule, and Related
  Spectroscopies: Science and Applications (HBSM 2009) - Issue dedicated to
  Ivan Lorgere and Oliver Guillot-Noel}\BibitemShut {NoStop}%
\bibitem [{\citenamefont {Sun}\ and\ \citenamefont {Waks}(2016)}]{Waks_2016}%
  \BibitemOpen
  \bibfield  {author} {\bibinfo {author} {\bibfnamefont {S.}~\bibnamefont
  {Sun}}\ and\ \bibinfo {author} {\bibfnamefont {E.}~\bibnamefont {Waks}},\
  }\href@noop {} {\bibfield  {journal} {\bibinfo  {journal} {arXiv:}\ }\textbf
  {\bibinfo {volume} {1602.04367}} (\bibinfo {year} {2016})}\BibitemShut
  {NoStop}%
\bibitem [{\citenamefont {Siyushev}\ \emph {et~al.}(2014)\citenamefont
  {Siyushev}, \citenamefont {Xia}, \citenamefont {Reuter}, \citenamefont
  {Jamali}, \citenamefont {Zhao}, \citenamefont {Yang}, \citenamefont {Duan},
  \citenamefont {Kukharchyk}, \citenamefont {Wieck}, \citenamefont {Kolesov},\
  and\ \citenamefont {Wratchtrup}}]{Kolesov_2014}%
  \BibitemOpen
  \bibfield  {author} {\bibinfo {author} {\bibfnamefont {P.}~\bibnamefont
  {Siyushev}}, \bibinfo {author} {\bibfnamefont {K.}~\bibnamefont {Xia}},
  \bibinfo {author} {\bibfnamefont {R.}~\bibnamefont {Reuter}}, \bibinfo
  {author} {\bibfnamefont {M.}~\bibnamefont {Jamali}}, \bibinfo {author}
  {\bibfnamefont {N.}~\bibnamefont {Zhao}}, \bibinfo {author} {\bibfnamefont
  {N.}~\bibnamefont {Yang}}, \bibinfo {author} {\bibfnamefont {C.}~\bibnamefont
  {Duan}}, \bibinfo {author} {\bibfnamefont {N.}~\bibnamefont {Kukharchyk}},
  \bibinfo {author} {\bibfnamefont {A.}~\bibnamefont {Wieck}}, \bibinfo
  {author} {\bibfnamefont {R.}~\bibnamefont {Kolesov}}, \ and\ \bibinfo
  {author} {\bibfnamefont {J.}~\bibnamefont {Wratchtrup}},\ }\href@noop {}
  {\bibfield  {journal} {\bibinfo  {journal} {Nat. Comm.}\ }\textbf {\bibinfo
  {volume} {5}},\ \bibinfo {pages} {3895} (\bibinfo {year} {2014})}\BibitemShut
  {NoStop}%
\bibitem [{\citenamefont {Probst}\ \emph {et~al.}(2014)\citenamefont {Probst},
  \citenamefont {Kukharchyk}, \citenamefont {Rotzinger}, \citenamefont
  {Tkal$\check{c}$ec}, \citenamefont {W\"{u}nsch}, \citenamefont {Wieck},
  \citenamefont {Siegel}, \citenamefont {Ustinov},\ and\ \citenamefont
  {Bushev}}]{Bushev_2014}%
  \BibitemOpen
  \bibfield  {author} {\bibinfo {author} {\bibfnamefont {S.}~\bibnamefont
  {Probst}}, \bibinfo {author} {\bibfnamefont {N.}~\bibnamefont {Kukharchyk}},
  \bibinfo {author} {\bibfnamefont {H.}~\bibnamefont {Rotzinger}}, \bibinfo
  {author} {\bibfnamefont {A.}~\bibnamefont {Tkal$\check{c}$ec}}, \bibinfo
  {author} {\bibfnamefont {S.}~\bibnamefont {W\"{u}nsch}}, \bibinfo {author}
  {\bibfnamefont {A.~D.}\ \bibnamefont {Wieck}}, \bibinfo {author}
  {\bibfnamefont {M.}~\bibnamefont {Siegel}}, \bibinfo {author} {\bibfnamefont
  {A.~V.}\ \bibnamefont {Ustinov}}, \ and\ \bibinfo {author} {\bibfnamefont
  {P.~A.}\ \bibnamefont {Bushev}},\ }\href {\doibase
  http://dx.doi.org/10.1063/1.4898696} {\bibfield  {journal} {\bibinfo
  {journal} {Applied Physics Letters}\ }\textbf {\bibinfo {volume} {105}},\
  \bibinfo {pages} {162404} (\bibinfo {year} {2014})}\BibitemShut {NoStop}%
\bibitem [{\citenamefont {Thiel}\ \emph {et~al.}(2011)\citenamefont {Thiel},
  \citenamefont {B\"ottger},\ and\ \citenamefont {Cone}}]{Cone_2011}%
  \BibitemOpen
  \bibfield  {author} {\bibinfo {author} {\bibfnamefont {C.}~\bibnamefont
  {Thiel}}, \bibinfo {author} {\bibfnamefont {T.}~\bibnamefont {B\"ottger}}, \
  and\ \bibinfo {author} {\bibfnamefont {R.~L.}\ \bibnamefont {Cone}},\
  }\href@noop {} {\bibfield  {journal} {\bibinfo  {journal} {J. Luminescence}\
  }\textbf {\bibinfo {volume} {131}},\ \bibinfo {pages} {353} (\bibinfo {year}
  {2011})}\BibitemShut {NoStop}%
\bibitem [{\citenamefont {Wolfowicz}\ \emph {et~al.}(2015)\citenamefont
  {Wolfowicz}, \citenamefont {Maier-Flaig}, \citenamefont {Marino},
  \citenamefont {Ferrier}, \citenamefont {Vezin}, \citenamefont {Morton},\ and\
  \citenamefont {Goldner}}]{Goldner_2015}%
  \BibitemOpen
  \bibfield  {author} {\bibinfo {author} {\bibfnamefont {G.}~\bibnamefont
  {Wolfowicz}}, \bibinfo {author} {\bibfnamefont {H.}~\bibnamefont
  {Maier-Flaig}}, \bibinfo {author} {\bibfnamefont {R.}~\bibnamefont {Marino}},
  \bibinfo {author} {\bibfnamefont {A.}~\bibnamefont {Ferrier}}, \bibinfo
  {author} {\bibfnamefont {H.}~\bibnamefont {Vezin}}, \bibinfo {author}
  {\bibfnamefont {J.~J.~L.}\ \bibnamefont {Morton}}, \ and\ \bibinfo {author}
  {\bibfnamefont {P.}~\bibnamefont {Goldner}},\ }\href {\doibase
  10.1103/PhysRevLett.114.170503} {\bibfield  {journal} {\bibinfo  {journal}
  {Phys. Rev. Lett.}\ }\textbf {\bibinfo {volume} {114}},\ \bibinfo {pages}
  {170503} (\bibinfo {year} {2015})}\BibitemShut {NoStop}%
\end{thebibliography}%

\end{document}